\documentclass[a4paper,twocolumn,amsmath,amssymb,superscriptaddress]{revtex4}
\usepackage{graphicx}

\newcommand{\traceSub}[2]{\textrm{Tr}_{#1} \! \left(#2\right)}

\newcommand{\abs}[1]{\left|#1\right|}

\newcommand{\integ}[1]{\ensuremath{\int \!\! \mathrm{d}#1 \,}}
\newcommand{\integlim}[3]{\ensuremath{\int_{#1}^{#2} \!\!\! \mathrm{d}#3 \,}}

\newcommand{\deriv}[2]{\frac{\textrm{d} #1}{\textrm{d} #2}}

\newcommand{\mean}[1]{\ensuremath{\left\langle #1 \right\rangle}}

\newcommand{\var}[1]{\ensuremath{\sigma_{#1}^2 }}

\newcommand{\prob}[1]{\textrm{Pr} \! \left(#1\right)}

\newcommand{\N}{\ensuremath{\mathcal{N}}}

\newcommand{\nbar}{\ensuremath{\bar{n}}}
\newcommand{\nbareff}{\ensuremath{\bar{n}_{\textrm{eff}}}}

\newcommand{\xzp}{\ensuremath{x_0}}

\newcommand{\omegam}{\ensuremath{\omega_M}}
\newcommand{\gammam}{\ensuremath{\gamma_M}}

\renewcommand{\H}{\ensuremath{H}}

\newcommand{\bra}[1]{\ensuremath{\left\langle #1 \right|}}
\newcommand{\ket}[1]{\ensuremath{\left| #1 \right\rangle}}

\newcommand{\Ups}{\ensuremath{\Upsilon}}
\newcommand{\Upsd}{\ensuremath{\Upsilon^\dagger}}

\newcommand{\D}{\ensuremath{D}}
\newcommand{\Sq}{\ensuremath{S}}

\renewcommand{\a}{\ensuremath{a}}
\renewcommand{\b}{\ensuremath{b}}

\newcommand{\ad}{\ensuremath{a^\dagger}}
\newcommand{\bd}{\ensuremath{b^\dagger}}

\newcommand{\ta}{\ensuremath{\tilde{a}}}

\newcommand{\alphai}{\ensuremath{\alpha_{\textrm{in}}}}

\newcommand{\ai}{\ensuremath{a^{\phantom{\dagger}}_{\textrm{in}}}}

\newcommand{\tai}{\ensuremath{\tilde{a}^{\phantom{\dagger}}_{\textrm{in}}}}

\newcommand{\tao}{\ensuremath{\tilde{a}^{\phantom{\dagger}}_{\textrm{out}}}}

\newcommand{\alphalo}{\alpha_{\textrm{LO}}}

\newcommand{\XM}{\ensuremath{X_M}}
\newcommand{\PM}{\ensuremath{P_M}}
\newcommand{\XMpar}[1]{\ensuremath{X_M^{#1}}}

\newcommand{\XMin}{\ensuremath{X_M^{\textrm{in}}}}

\newcommand{\XMBar}{\ensuremath{\overline{X}_M}}
\newcommand{\PMBar}{\ensuremath{\overline{P}_M}}
\newcommand{\XMBarin}{\ensuremath{\overline{X}_M^{\textrm{\,in}}}}
\newcommand{\XMBarout}{\ensuremath{\overline{X}_M^{\textrm{\,out}}}}
\newcommand{\PMBarin}{\ensuremath{\overline{P}_M^{\textrm{\,in}}}}
\newcommand{\PMBarout}{\ensuremath{\overline{P}_M^{\textrm{\,out}}}}

\newcommand{\XMTilde}{\ensuremath{\widetilde{X}_M}}

\newcommand{\XL}{\ensuremath{X_L}}
\newcommand{\PL}{\ensuremath{P_L}}

\newcommand{\XC}[1]{\ensuremath{X_{C #1}}}

\newcommand{\PLm}{\ensuremath{P_L}}
\newcommand{\PLmFirst}{\ensuremath{P^{(1)}_L}}
\newcommand{\PLmSecond}{\ensuremath{P^{(2)}_L}}

\newcommand{\PLout}{\ensuremath{P_L^{\textrm{out}}}}
\newcommand{\PLin}{\ensuremath{P_L^{\textrm{in}}}}

\newcommand{\etal}{\emph{et al}.~}

\begin{document}

\title{Pulsed quantum optomechanics}\thanks{This work has been published in Proc Natl Acad Sci USA \textbf{108}, 16182 (2011). [Open access article.]}

\author{M.~R.~Vanner}\email[E-mail: ]{michael.vanner@univie.ac.at}
\affiliation{Vienna Center for Quantum Science and Technology (VCQ),
Faculty of Physics, University of Vienna, Boltzmanngasse 5, Vienna A-1090, Austria}

\author{I.~Pikovski}
\affiliation{Vienna Center for Quantum Science and Technology (VCQ),
Faculty of Physics, University of Vienna, Boltzmanngasse 5, Vienna A-1090, Austria}

\author{G.~D.~Cole}
\affiliation{Vienna Center for Quantum Science and Technology (VCQ),
Faculty of Physics, University of Vienna, Boltzmanngasse 5, Vienna A-1090, Austria}

\author{M.~S.~Kim}
\affiliation{QOLS, Blackett Laboratory, Imperial College London, London SW7 2BW, United Kingdom}

\author{\v{C}.~Brukner}
\affiliation{Vienna Center for Quantum Science and Technology (VCQ),
Faculty of Physics, University of Vienna, Boltzmanngasse 5, Vienna A-1090, Austria}
\affiliation{Institute for Quantum Optics and Quantum Information (IQOQI) of the
Austrian Academy of Sciences, A-1090 Vienna and A-6020 Innsbruck, Austria}

\author{K.~Hammerer}
\affiliation{Institute for Quantum Optics and Quantum Information (IQOQI) of the
Austrian Academy of Sciences, A-1090 Vienna and A-6020 Innsbruck, Austria}
\affiliation{{Institute for Theoretical Physics and Albert Einstein Institute, University of Hannover, Callinstr. 38, D-30167 Hannover, Germany}}

\author{G.~J.~Milburn}
\affiliation{{School of Mathematics and Physics, The University of Queensland, Australia 4072}}

\author{M.~Aspelmeyer}
\affiliation{Vienna Center for Quantum Science and Technology (VCQ),
Faculty of Physics, University of Vienna, Boltzmanngasse 5, Vienna A-1090, Austria}

\begin{abstract}
Studying mechanical resonators via radiation pressure offers a rich avenue for the exploration of quantum mechanical behavior in a macroscopic regime. However, quantum state preparation and especially quantum state reconstruction of mechanical oscillators remains a significant challenge. Here we propose a scheme to realize quantum state tomography, squeezing and state purification of a mechanical resonator using short optical pulses. The scheme presented allows observation of mechanical quantum features despite preparation from a thermal state and is shown to be experimentally feasible using optical microcavities. Our framework thus provides a promising means to explore the quantum nature of massive mechanical oscillators and can be applied to other systems such as trapped ions.
\end{abstract}

\maketitle

%%%%********************************************************************************%%%
\section{Introduction}

Coherent quantum mechanical phenomena, such as entanglement and superposition, are not apparent in the macroscopic realm. It is currently held that on large scales quantum mechanical behavior is masked by decoherence~\cite{ref:decoherence} or that quantum mechanical laws may even require modification~\cite{ref:collapse}. Despite substantial experimental advances, see for example Ref.~\cite{ref:molecule}, probing this regime remains extremely challenging. Recently however, it has been proposed to utilize the precision and control of quantum optical fields in order to investigate the quantum nature of massive mechanical resonators by means of the radiation pressure interaction ~\cite{ref:BJK, ref:revivals, ref:review}. Quantum state preparation and the ability to probe the dynamics of mechanical oscillators, the most rigorous method being quantum state tomography, are essential for such investigations. These have been experimentally realized for various quantum systems, e.g. light~\cite{ref:Smithey1993, ref:lightReview}, trapped ions \cite{ref:ions},  atomic ensemble spin \cite{ref:spinSqueeze} and intra-cavity microwave fields~\cite{Deleglise2008}. By contrast, an experiment realizing both quantum state preparation and tomography of a mechanical resonator is yet to be achieved. Also, schemes that can probe quantum features without full tomography (e.g.~\cite{ref:Armour2002, ref:revivals}) are similarly challenging. In nano-electromechanics, cooling of resonator motion and preparation of the ground state has been observed \cite{ref:Rocheleau2010, ref:OConnell2010} but quantum state reconstruction \cite{Rabl2004} remains outstanding. In cavity optomechanics significant experimental progress has been made towards quantum state control over mechanical resonators \cite{ref:review}, which includes classical phase-space monitoring \cite{ref:phaseSpace}, laser cooling close to the ground state \cite{ref:cooling} and low noise continuous measurement of mechanically induced phase fluctuations \cite{ref:imprecision}. Still, quantum state preparation is technically difficult primarily due to thermal bath coupling and weak radiation pressure interaction strength, and quantum state reconstruction remains little explored. Thus far, a common theme in proposals for mechanical state reconstruction is state transfer to and then readout of an auxillary quantum system~\cite{ref:QSR}. This is a technically demanding approach and remains a challenge.

In this paper we introduce an optomechanical scheme that provides direct access to all the mechanical quadratures in order to obtain full knowledge about the quantum state of mechanical motion. This is achieved by observing the distribution of phase noise of strong pulses of light at various times throughout a mechanical period. We show that the same experimental tools used for quantum state tomography can also be used for squeezed state preparation and state purification, which thus provides a complete experimental framework. Our scheme does not require `cooling via damping' \cite{ref:review} and can be performed within a single mechanical cycle thus significantly relaxing the technical requirements to minimize thermal contributions from the environment.

\begin{figure}[t]
\includegraphics[width=0.9\hsize]{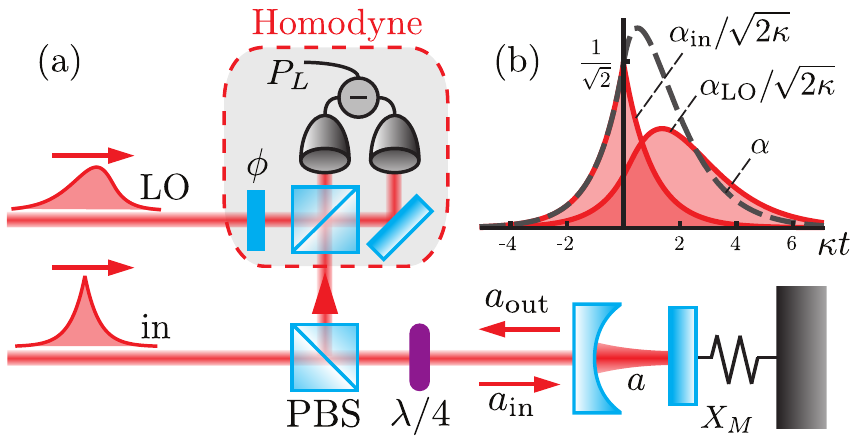}%\vspace{-2mm}
\caption{(a) Schematic of the optical setup to achieve measurement based quantum state engineering and quantum state tomography of a mechanical resonator. An incident pulse (in) resonantly drives an optomechanical cavity, where the intracavity field $a$ accumulates phase with the position quadrature $\XM$ of a mechanical oscillator. The field emerges from the cavity (out) and balanced homodyne detection is used to measure the optical phase with a local oscillator pulse (LO) shaped to maximize the measure of the mechanical position. (b) Scaled envelopes of the optimal input pulse, its corresponding intracavity field and the optimal local oscillator as computed in the appendix.}
\label{Fig:Setup}
\end{figure}

Using a pulsed interaction that is very short compared to the period of an oscillator to provide a back-action-evading measurement of position was introduced in the seminal contributions of Braginsky and colleagues~\cite{ref:Braginsky1978, ref:BraginskyBook}, where schemes for sensitive force detection were developed. In our work, the quantum nature of a mechanical resonator is itself the central object of investigation. Here, the pulsed interaction is used to provide an experimentally feasible means to generate and fully reconstruct quantum states of mechanical motion. The proposed experimental setup is shown in Fig.~\ref{Fig:Setup}. A pulse of duration much less than the mechanical period is incident upon an optomechanical Fabry-P\'{e}rot cavity which we model as being single-sided. Due to the entanglement generated during the radiation-pressure interaction, the optical phase becomes correlated with the mechanical position while the optical intensity imparts momentum to the mechanical oscillator. Time-domain homodyne detection \cite{ref:lightReview} is then used to determine the phase of the field emerging from the cavity, and thus to obtain a measure of the mechanical position. For each pulse, the measurement outcome $\PLm$ is recorded, which for Gaussian optical states has mean and variance
\begin{equation}
\label{eq:meanVarPL}
\mean{\PL} = \chi\mean{\XMin}, \qquad \var{\PL} = \var{\PLin} + \chi^2 \var{\XMin},
\end{equation}
respectively. $\XMin$ is the mechanical position quadrature immediately prior to the interaction and $\PLin$ describes the input phase of light. The position measurement strength $\chi$ is an important parameter in this work as it quantifies the scaling of the mechanical position information onto the light field. A derivation of Eq. (\ref{eq:meanVarPL}) including an optimization of $\chi$ by determining the input pulse envelope to gain the largest cavity enhancement is provided in the appendix.

In order to describe and quantify the pulse interaction and measurement we use the non-unitary operator $\Ups$ that determines the new mechanical state via $\rho^{\textrm{out}}_M \propto \Ups \rho^{\textrm{in}}_M \Upsd$. This operator is mechanical state independent and can be determined from the probability density of measurement outcomes 

\begin{equation}
\label{eq:propPL}
\prob{\PLm} = \traceSub{M}{\Upsd \Ups \rho_M^{\textrm{in}}}.
\end{equation}
For pure optical input, it takes the form
\begin{equation}
\Ups = ( \pi 2 \var{\PLin} )^{-\frac{1}{4}} \, \exp\left[i\Omega \XM - \frac{ \left(\PLm - \chi \XM \right)^2}{4 \var{\PLin}}\right],
\label{eq:Upsilon}
\end{equation}
where $\Omega$ quantifies the momentum transfer to the mechanics due to the pulse mean photon number. $\Ups$ can be readily understood by considering its action on a mechanical position wavefunction. It selectively narrows the wavefunction to a width scaling with $\chi^{-2}$ about a position which depends upon the measurement outcome. Moreover, the quantum non-demolition-like nature of $\Ups$ allows for back-action-evading measurements of $\XM$, i.e. the back-action noise imparted by the quantum measurement process occurs in the momentum quadrature only \footnote{This is because our measurement operator commutes with the mechanical position, since the mechanical evolution can be neglected during the short optomechanical interaction.}. Other methods, such as the continuous variational measurement scheme~\cite{Vyatchanin1993}, which has recently been considered for gravitational-wave detectors~\cite{ChenConditional}, also allow for back-action-evading measurements. However, using short pulses offers a technically simpler route for quantum state tomography and is readily implementable, as will be discussed below.

In the following, we consider coherent drive i.e. $\var{\PLin} = 1/2$. We first address the important challenge of how to experimentally determine the motional quantum state of a mechanical resonator. We then discuss how such a measurement can be used for quantum state preparation and finally we provide details for a physical implementation and analyze a thorough list of potential experimental limitations.

%%%%********************************************************************************%%%
\section{Mechanical quantum state tomography}

Of vital importance to any experiment aiming to explore quantum mechanical phenomena is a means to measure coherences and complementary properties of the quantum system. This is best achieved by complete quantum state tomography, which despite being an important quantum optical tool has recieved very little attention for mechanical resonators~\footnote{During the submission process of this manuscript a scheme to perform tomography of the motional state of a trapped particle using a time-of-flight expansion was proposed~\cite{Romero2010}.}. Any measurement made on a single realization of a quantum state cannot yield sufficient information to completely characterize that quantum state. The essence of quantum state tomography is to make measurements of a specific set of observables over an ensemble of identically prepared realizations. The set is such that the measurement results provide sufficient information for the quantum state to be uniquely determined. One such method is to measure the  marginals $\bra{X} e^{-i\theta n} \rho e^{i\theta n} \ket{X}$, where $n$ is the number operator, for all phase-space angles $\theta$, see Refs.~\cite{ref:Smithey1993, ref:lightReview, Vogel1989} and e.g. Ref. \cite{ref:Dunn1995}.

Our scheme provides a means for precision measurement of the mechanical quadrature marginals, thus allowing the mechanical quantum state to be determined. Specifically, given a mechanical state $\rho_M^{\textrm{in}}$, harmonic evolution of angle $\theta = \omegam t$ provides access to all the quadratures of this mechanical quantum state which can then be measured by a subsequent pulse. Thus, reconstruction of any mechanical quantum state can be performed. The optical phase distribution (\ref{eq:propPL}), including this harmonic evolution, becomes
\begin{multline}
\prob{\PLm} = \\
 \int \frac{\textrm{d}\XM}{\sqrt{\pi}} e^{- \left(\PLm - \chi \XM \right)^2} \bra{\XM} e^{-i\theta n} \rho_M^{\textrm{in}} e^{i\theta n} \ket{\XM}, 
\label{eq:probPLconv}
\end{multline}
which is a convolution between the mechanical marginal of interest and a kernel that is dependent upon $\chi$ and the quantum phase noise of light. The effect of the convolution is to broaden the marginals and to smooth any features present.

\begin{figure}[]
\includegraphics[]{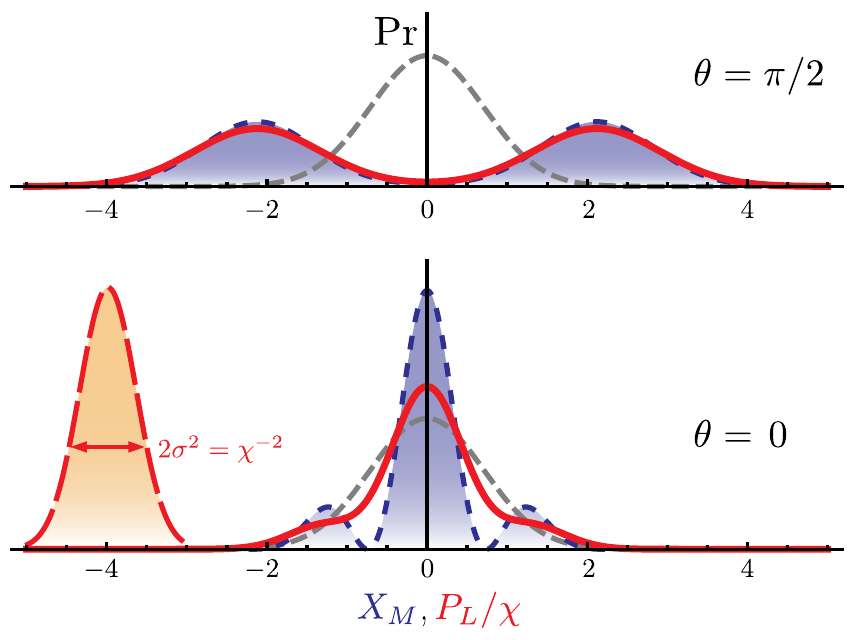}
\caption{The scheme presented here provides an experimentally feasible means to obtain direct access to the marginals of a quantum state of a mechanical resonator. Shown are complementary quadrature marginals of the mechanical coherent state superposition $\ket{\psi_\delta} \propto \ket{i \delta} + \ket{-i \delta}$, for $\delta = 1.5$ (blue dashed lines with fill, plotted with $\XM$). The mechanical ground state is shown for comparison in gray dashed lines. The two population components are seen for the quadrature angle $\theta = \pi/2$ and the quantum interference fringes for $\theta = 0$. A coherent optical pulse is used to probe the mechanical state where its phase quadrature becomes the convolution between the intrinsic phase noise, with variance scaling with $\chi^{-2}$, and the mechanical marginal (red solid lines, plotted with $\PL/\chi$ where $\chi = 2$), see Eq. (\ref{eq:probPLconv}). The convolution kernel can be observed by using a fixed length cavity, shown in the $\theta = 0$ plot (red dashed line with fill, fixed length with $\XM = -4$), which allows for accurate recovery of the mechanical marginals even for a weak measurement strength $\chi$.}
\label{Fig:ConvMarginals}
\end{figure}

Let us consider the specific example of a mechanical resonator in a superposition of two coherent states, i.e. $\ket{\psi_\delta} \propto \ket{i \delta} + \ket{-i \delta}$. The $\XM$ marginal of this mechanical Schr\"{o}dinger-cat state contains oscillations on a scale smaller than the ground state. The convolution scales the amplitude of these oscillations by $\exp(-\frac{2 \delta^2}{\chi^2 + 1})$ and thus for small $\chi$ they become difficult to resolve  in the optical phase noise distribution. Shown in Fig.~\ref{Fig:ConvMarginals} are marginals of the mechanical state $\ket{\psi_\delta}$ and the optical phase distributions that would be observed according to (\ref{eq:probPLconv}). Scaling the phase distribution by using the variable $\PLm/\chi$ provides an approximation to the mechanical marginals, which becomes more accurate with increasing $\chi$ and may even show the interference features in a superposition state. Indeed, the limiting case of infinite $\chi$ corresponds to a von-Neumann projective measurement of the mechanical position, such that the distribution obtained for $\PLm/\chi$ becomes identical to the mechanical marginals. However, \emph{the mechanical marginals can be recovered even for small measurement strength $\chi$}. This is achieved as follows: First, by fixing the length of the cavity the optical phase distribution can be observed without contributions from mechanical position fluctuations. This allows measurement of the convolution kernel for a particular $\chi$ (determined by the properties of the mechanical resonator of interest, cavity geometry and pulse strength, see (\ref{eq:maxChi})). With $\chi$ and the kernel known one can then perform deconvolution to determine the mechanical marginals. The performance of such a deconvolution is limited by experimental noise in the calibration of $\chi$ and the measurement of $\prob{\PLm}$. However, it is expected that these can be accurately measured as quantum noise limited detection is readily achieved.

%%%%********************************************************************************%%%
\section{Mechanical quantum state engineering and characterization}

We now discuss how the measurement affects the mechanical state. First, we consider  $\Ups$ acting on a mechanical coherent state $\ket{\beta}$. By casting the exponent of $\Ups$ in a normal ordered form, one can show that the resulting  mechanical state, which is conditioned on measurement outcome $\PLm$, is $\N_{\beta} \Ups \ket{\beta} =  \Sq(r) \D(\mu_{\beta}) \ket{0}$. Here, $\N_{\beta}$ is a $\beta$-dependent normalization, $\D$ is the displacement operator for $\mu_{\beta} = (\sqrt{2} \beta + i \Omega +  \chi \PLm ) / \sqrt{2(\chi^2 + 1)}$ and $\Sq$ is the squeezing operator, which yields the position width $2\var{\XM} = e^{-2r} = (\chi^2 + 1)^{-1}$.

In most experimental situations, the initial mechanical state is in a thermal state $\rho_{\nbar}= \frac{1}{\pi \nbar} \int \! \textrm{d}^2 \! \beta e^{-|\beta|^2/\nbar} \ket{\beta}\bra{\beta}$, quantified by its average phonon occupation number $\nbar$.  The marginals of the resulting state after the action of $\Ups$ are
\begin{multline}
\label{eq:KimmThermalState}
\bra{\XM} e^{-i\theta n} \Ups \, \rho_{\nbar}  \Ups^{\dagger} e^{i\theta n} \ket{\XM}
 \propto \\
 \exp \left[-\frac{(\XM - \mean{\XMpar{\theta}}  )^2}{2 \sigma^2_{\theta}} \right],
\end{multline}
where 
\begin{equation}
\label{eq:KimmThermal}
\begin{split}
\mean{\XMpar{\theta}} & =   \frac{ \chi \, \PLm}{\chi^2 + \frac{1}{1+ 2 \bar{n}}} \, \cos(\theta) - \Omega \, \sin(\theta) ,\\ 
\\
\sigma^2_{\theta} & = \frac{1}{2} \frac{\cos^2(\theta)}{ \chi^2 + \frac{1}{1+ 2 \bar{n}}}  + \frac{1}{2}(\chi^2 +1 + 2 \bar{n}) \, \sin^2(\theta)
\end{split}
\end{equation}
are the mean and variance of the resulting conditional state, respectively. For large initial occupation (provided thermal fluctuations are negligible during the short interaction), the resultant position quadrature of the mechanics has mean $\mean{\XMpar{\theta=0}} \simeq  \PLm / \chi$ and width $2 \sigma^2_{\theta=0} \simeq \chi^{-2}$. Thus, squeezing in the $\XM$ quadrature below the ground state is obtained when $\chi > 1$ and is \emph{independent of the initial thermal occupation of the mechanics.} We have thus shown how the remarkable behavior of quantum measurement (also used in Refs.~\cite{ref:spinSqueeze,Deleglise2008, ref:opticalQND, Hammerer2009}) can be experimentally applied to a mechanical resonator for quantum state preparation.

\begin{figure}[]
\includegraphics[]{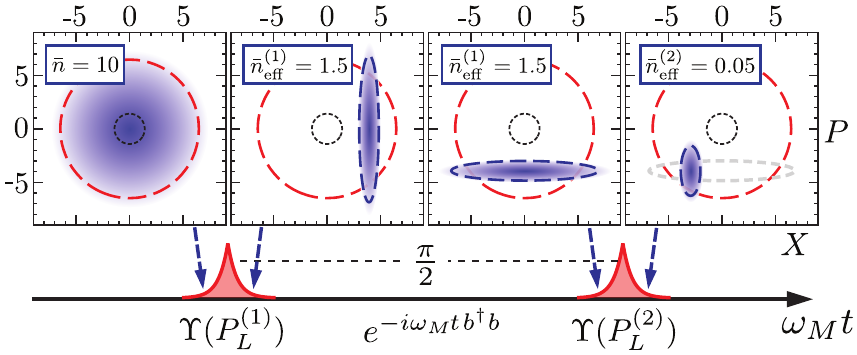}%\vspace{-2mm}
\caption{Wigner functions of the mechanical state (above) at different times (indicated by arrows) during the experimental protocol (below). From left: Starting with an initial thermal state $\nbar = 10$, (this is chosen to ensure the figure dimensions are reasonable,) a pulsed measurement is made with $\chi = 1.5$ and outcome $\PLmFirst = 4 \chi$ obtained, which yields an $\XM$ quadrature squeezed state. The mechanical state evolves into a $\PM$ quadrature squeezed state following free harmonic evolution of one quarter of a mechanical period prior to a second pulse with outcome $\PLmSecond = -3\chi$ yielding the high purity mechanical squeezed state. The effective thermal occupation of the mechanical states during the protocol is indicated. The final state's occupation can be reduced below unity even for large initial occupation, see Eq.~(\ref{eq:nbarEff2}) of the main text. Dashed lines indicate the $2\sigma$-widths and the dotted lines show the ground state ($\nbar = 0$) for comparative purposes. The displacement $\Omega$ is not shown.}
\label{Fig:Squeeze}
\end{figure}

There is currently significant interest in the preparation of low entropy states of mechanical resonators as a starting point for quantum experiments, e.g. Refs. \cite{ref:Rocheleau2010, ref:OConnell2010, ref:cooling}. The two main methods being pursued in optomechanics \cite{ref:review} are `passive cooling' which requires the stable operation of a (usually cryogenically compatible) high-finesse cavity, and `active cooling' which requires precision measurement and feedback. Closer in spirit to the latter, our pulsed measurement scheme provides a third method to realize high-purity states of the mechanical resonator. We quantify the state purity after measurement via an effective mechanical thermal occupation $\nbareff$, which we define through $1 + 2 \nbareff = \sqrt{4 \var{\theta=0} \var{\theta=\pi/2}}$. When acting on an initial thermal state, the measurement dramatically reduces uncertainty in the $\XM$ quadrature, but leaves the thermal noise in the $\PM$ quadrature unchanged: use of (\ref{eq:KimmThermal}) for $\nbar \gg 1$ yields $\bar{n}_{\textrm{eff}}^{(1)} \approx \sqrt{\bar{n} / 2 \chi^2}$. The purity can be further improved by a second pulse, which is maximized for pulse separation $\theta = \omegam t = \pi/2$, where the initial uncertainty in the momentum becomes the uncertainty in position. Such a sequence of pulses\footnote{We note that strong squeezing of an oscillator can also be achieved by using rapid modifications to the potential at quarter period intervals \cite{Janszky1992}. However, we would like to emphasize that the squeezing we are discussing here does not arise from a parametric process, see e.g. Ref. \cite{Walls1983}, rather it is due to the non-unitary action of measurement.} is represented in Fig.~\ref{Fig:Squeeze}, where the resulting state was obtained akin to (\ref{eq:KimmThermalState}). The effective occupation of the final state after two pulses  is
\begin{equation}
\nbareff^{(2)}  \simeq \frac{1}{2} \left( \sqrt{1 + \frac{1}{\chi^4}} - 1 \right),
\label{eq:nbarEff2}
\end{equation}
which is also independent of initial occupation. For $\chi > 1$, $\nbareff^{(2)}$ is well below unity and therefore this scheme can be used as an alternative to `cooling via damping' for mechanical state purification.

Following state preparation, one can use a subsequent `read-out' pulse after an angle of mechanical free evolution $\theta$ to perform tomography. During state preparation however, the random measurement outcomes will result in random mechanical means (\ref{eq:KimmThermal}). This can be overcome by recording and utilizing the measurement outcomes. One can achieve unconditional state preparation with use of appropriate displacement prior to the read-out pulse. Or, use post-selection to analyze states prepared within a certain window. Alternatively, one may compensate during data analysis by appropriately adjusting each measurement outcome obtained during read-out. We now look at the latter option and consider a Gaussian mechanical state prepared by a prior pulsed measurement. The position distribution has variance $\sigma^2$ to be characterized and has a known mean $\mean{\XMpar{(p)}}$, which is dependent upon the random measurement outcome. The read-out pulse will then have the distribution $\prob{\PLm} \propto \exp \left[ (-(\PLm - \chi \mean{\XMpar{(p)}})^2)/(1 + \chi^2 2\sigma^2) \right]$. For each read-out pulse, by taking $\PLm |_p = \PLm - \chi\mean{X^{(p)}}$ one can obtain the conditional variance $\var{\PL |_p}$ for all $\theta$ to characterize the noise of the prepared Gaussian state. We note that this concept of compensating for a random but known mean can also be used to characterize non-Gaussian states.

%%%*********************************************************************************%%%
\section{Experimental feasibility}

\begin{figure}[]
\includegraphics[]{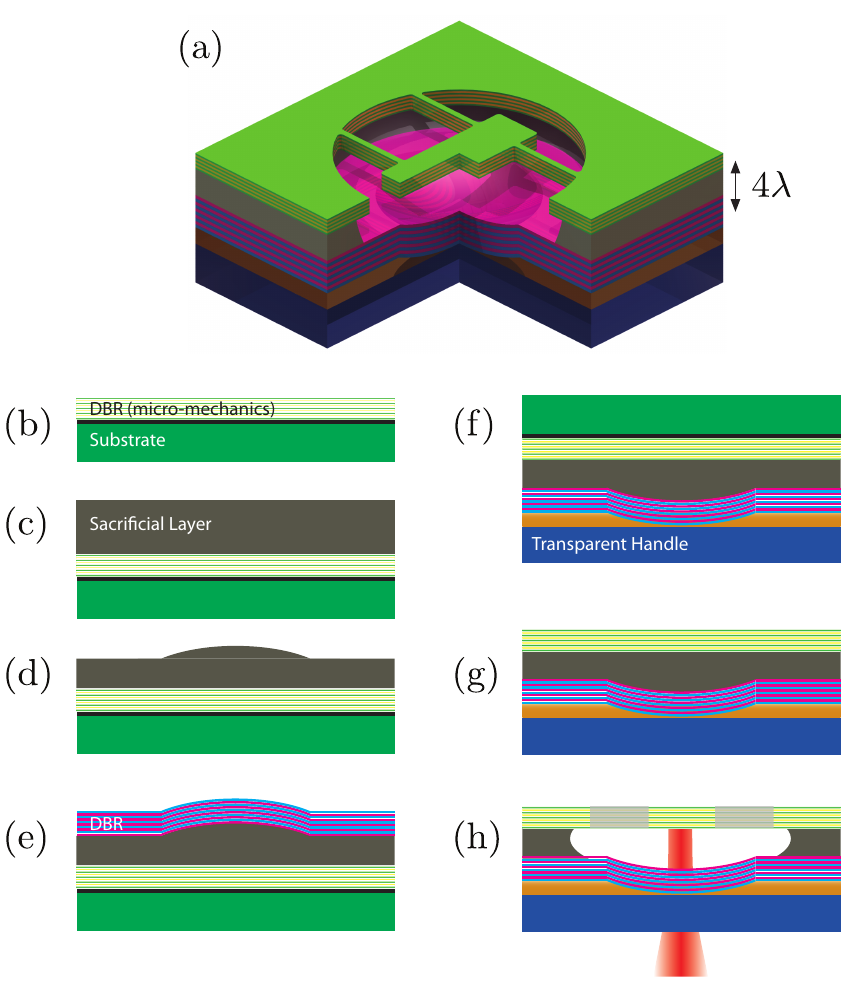}%\vspace{-2mm}
\caption{Our proposed design and fabrication procedure for high-finesse optomechanical microcavities: Using microcavities provides optomechanical coupling rates many orders of magnitude larger than current millimetre or centimetre length-scale implementations of optomechanical Fabry-P\'{e}rot cavities and can provide sufficient radiation-pressure interaction to resolve the small-scale quantum properties of the mechanical resonator. (a) Cross-sectional view with a quarter of the device removed. Uppermost (colored green) is the mechanical resonator supported by auxiliary beams as was considered in Ref.~\cite{ref:ColeNatComm}. The optical field is injected into the device from below through a transparent handle (colored blue) and the curved rigid input mirror (colored pink) and then resonates in the vacuum-gap between this and the mechanical device before being retro-reflected. The design is a layered structure, fabricated in the following steps: (b) The base consists of a high-reflectivity distributed Bragg reflector (DBR) and an etch stop layer deposited on a suitable handle substrate. (c) First, a sacrificial film is deposited atop the DBR. (d) Next, a microlens pattern is transferred into the sacrificial layer through a reflow and reactive ion etching process. The radius of curvature of this structure is designed to match the phase front of the optical mode to minimize diffraction loss. (e) Following the microlens fabrication process a high reflectivity dielectric DBR is deposited over the sample surface. (f) The structure is then flipped and bonded to a transparent handle using a suitable low-absorption adhesive (e.g. spin on glass or UV-curable epoxy). (g) After mounting, the original growth substrate and etch stop are removed via chemo-mechanical etching. (h) Finally, the mechanical resonator is patterned and subsequently released via selective removal of the underlying sacrificial film. We remark that these integrated structures provide a platform for `on-chip' hybridization with other quantum systems.}
\label{Fig:microCav}
\end{figure}

We now provide a route for experimental implementation, discussing potential limitations and an experimentally feasible parameter regime. To ensure that the interaction time be much less than mechanical time-scales the cavity decay rate $\kappa$ must be much larger than the mechanical frequency. To this end, we consider the use of optical microcavities operating at $\lambda = 1064$~nm, length $4\lambda$ and finesse of $7000$, which have an amplitude decay rate $\kappa/2\pi \simeq 2.5$~GHz. Such short cavity devices incorporating a micromechanical element as one of the cavity mirrors have previously been fabricated for tunable optical filters, vertical-cavity surface-emitting lasers (VCSELs) and amplifiers (see for example Ref.~\cite{ref:Cole2008}), but are yet to be considered for quantum optomechanical applications. Typically, these devices employ plane-parallel geometries, which places a severe constraint on the minimum lateral dimensions of the suspended mirror structure in order to minimize diffraction losses \cite{ref:Riazi1981}. Geometries using curved mirrors are required to reduce diffraction losses for the practical realization of high-finesse cavities. Presently, all realizations use a curved suspended mirror, see e.g. Ref. \cite{ref:curved}. However, in order to allow for enhanced freedom in the construction of the mechanical resonator, particularly with respect to the development of ultra-low loss mechanical devices \cite{ref:ColeNatComm}, a flat suspended mirror is desired. In Fig.~\ref{Fig:microCav} our proposed fabrication procedure for such a device is shown. The small-mode-volume cavity considered here provides the bandwidth necessary to accommodate the short optical pulses and additionally offers a large optomechanical coupling rate. One technical challenge associated with these microcavities is fabrication with sufficient tolerance to achieve the desired optical resonance (under the assumption of a limited range of working wavelength), however this can be overcome by incorporating electrically controlled tunability of the cavity length \cite{ref:Cole2008, ref:curved}.

For a mechanical resonator with eigenfrequency $\omega_M/2\pi = 500 $~kHz and effective mass $m = 10$~ng, the mechanical ground-state size is $\xzp = \sqrt{\hbar/ m\omegam} \simeq 1.8$~fm and optomechanical coupling proceeds at $g_0/2\pi = \omega_c(\xzp/\sqrt{2}L)/2\pi \simeq 86$~kHz, where $\omega_c$ is the mean cavity frequency and $L$ is the mean cavity length. The primary limitation in measurement strength is the optical intensity that can be homodyned before photodetection begins to saturate. Using pulses of mean photon number $N_p = 10^8$, which can be homodyned, yields $\Omega \simeq 10^{4}$ for the mean momentum transfer\footnote{This momentum is comparable to the width of a thermal state, i.e. $\Omega /\sqrt{\nbar} < 10$ for room temperature. Thus the mechanical motion remains harmonic. } and $\chi \simeq 1.5$. For this $\chi$, the action of a single pulse on a large thermal state reduces the mechanical variance to $\var{\XM} \simeq 0.2$, i.e. less than half the width of the ground-state. With a second pulse after mechanical evolution the effective occupation (\ref{eq:nbarEff2}) is $\nbareff^{(2)} \simeq 0.05$.

In order to observe mechanical squeezing, i.e. $\var{\XM}~<~1/2$, the conditional variance must satisfy $\var{\PL |_p} < \var{\PLin} + \chi^2/2$, where additional noise sources that do not affect the mechanical state, e.g. detector noise, can be subsumed into $\var{\PLin}$. It is therefore necessary to have an accurate experimental calibration of $\chi$ to quantify the mechanical width. (Similarly, $\Omega$ must also be accurately known to determine the conditional mean, see Eq.~(\ref{eq:KimmThermal}).) This can be performed in the laboratory as follows: For a fixed length cavity and a given pulse intensity, the length of the cavity is adjusted by a known amount (by a calibrated piezo for example) and the proportionality between the homodyne measurement outcomes and the cavity length is determined. The pulses are then applied to a mechanical resonator and $\chi$ is determined with knowledge of $\xzp$ of the resonator. With $\chi$ known $\Omega$ can then also be measured by observing the displacement of the mechanical state after one-quarter of a period.

Finally we discuss practical limitations. Firstly, finite mechanical evolution during the interaction decreases the back-action-evading nature of the measurement, which is described in the appendix. Such evolution is not expected to be a severe limitation in the proposed implementation considered here as $\omegam/\kappa \simeq 10^{-4}$. Secondly, the optical measurement efficiency $\eta$, affected by optical loss, inefficient detection and mode-mismatch, yields a reduced measurement strength $\chi \rightarrow \sqrt{\eta}\chi$. And thirdly, in many situations coupling to other mechanical vibrational modes is expected. This contributes to the measurement outcomes and yields a spurious broadening of the tomographic results for the mode of interest. In practice however, one can minimize these contributions by engineering mechanical devices with high effective masses for the undesired modes and tailoring the intensity profile of the optical spot to have good overlap with a particular vibrational profile~\cite{Pinard1999}.

%%%*********************************************************************************%%%
\section{Coupling to a thermal bath}
For our tomography scheme the mechanical quantum state must not be significantly perturbed during the time-scale $\omegam^{-1}$. To estimate the effect of the thermal bath following state preparation we consider weak and linear coupling to a Markovian bath of harmonic oscillators. For this model, assuming no initial correlations between the mechanics and the bath, the rethermalization scales with $\nbar \gammam$, where $\gammam$ is the mechanical damping rate. It follows that an initially squeezed variance $(\chi > 1)$ will increase to $1/2$ on a time-scale 
\begin{equation}
\tau = \frac{Q}{\nbar \omegam} \frac{1}{2}\left(1-\frac{1}{\chi^2} \right).
\end{equation}
Thus, for the parameters above and mechanical quality $Q = \omegam/\gammam \simeq 10^5$ a temperature $T \lesssim 1$~K is required for the observation of squeezing during one mechanical period.

The state purification protocol, as shown in Fig.~\ref{Fig:Squeeze}, is affected by rethermalization between the two pulsed measurements. This increases the effective thermal occupation and Eq.~(\ref{eq:nbarEff2}) is modified to
\begin{equation}
\nbareff^{(2)}(T)  \simeq \frac{1}{2} \left( \sqrt{1 + \frac{1}{\chi^4} + \frac{\pi \nbar}{Q \chi^2}} - 1 \right).
\end{equation}
For the above system parameters $\nbareff^{(2)}(T = 1~\textrm{K}) \simeq 0.15$. Thus, mechanical state purification by measurement is readily attainable even at a modest bath temperature. 

Moreover, we note that the position measurements of this scheme can be used to probe open system dynamics and thus provide an empirical means to explore decoherence and bath coupling models \cite{ref:BreuerBook}.

%%%%********************************************************************************%%%
\section{Conclusions}

We have described a scheme to overcome the current challenge of quantum state reconstruction of a mechanical resonator, which is of vital importance for the exploration of quantum mechanical phenomena of these macroscopic objects. Our experimental protocol allows for state purification, remote preparation of a mechanical squeezed state and direct measurements of the mechanical marginals for quantum state reconstruction, thus providing a complete experimental framework. The experimental feasibility has been analyzed and we have shown that with the use of optomechanical microcavities this scheme can be readily implemented. The optomechanical entanglement generated by the pulsed interaction may also be a useful resource for quantum information processing.  Moreover, the framework we have introduced can be built upon for further applications in quantum optomechanics and can be generalized to other systems, such as nano-electromechanics and superconducting resonators, or used with dispersive interaction to study the motional state of mechanical membranes, trapped ions or particles in a cavity.

%%%%********************************************************************************%%%
\appendix 

\section{Model}
\label{app:Model}

The intracavity optomechanical Hamiltonian in the rotating frame at the cavity frequency is 
\begin{equation}
\H = \hbar \omegam \bd\b - \hbar g_0 \ad\a (\b + \bd),
\end{equation}
where $\a$ ($\b$) is the optical (mechanical) field operator. The cavity field accumulates phase in proportion to the mechanical position and is driven by resonant radiation via the equation of motion
\begin{equation}
\label{eq:optLang}
\deriv{\a}{t} = ig_0(\b + \bd)a - \kappa a + \sqrt{2\kappa} \ai,
\end{equation}
where $\kappa$ is the cavity decay rate and $\ai$ describes the optical input including drive and vacuum. During a pulsed interaction of timescale $\kappa^{-1} \ll \omegam^{-1}$ the mechanical position is approximately constant. This decouples (\ref{eq:optLang}) from the corresponding mechanical equation of motion and during the short interaction we have $\deriv{\b}{t} \simeq ig_0\ad a$, where we neglect the mechanical harmonic motion, mechanical damping and noise processes. We write $\ai(t) =  \sqrt{N_p} \alphai(t) + \tai(t)$, where $\alphai(t)$ is the slowly varying envelope of the drive amplitude with $\integ{t} \alphai^2 = 1$ and $N_p$ is the mean photon number per pulse and similarly $a = \sqrt{N_p} \alpha(t) + \ta(t)$. Neglecting $ig_0(\b + \bd)\ta$ and approximating $\alpha$ as real, (\ref{eq:optLang}) becomes the pair of linear equations: 
\begin{eqnarray}
\deriv{\alpha}{t} = \sqrt{2\kappa} \, \alphai - \kappa \alpha \label{eq:dalphadt},\\
\deriv{\ta}{t} = i g_0 \sqrt{N_p} (\b + \bd) \alpha  + \sqrt{2\kappa} \, \tai - \kappa \ta . \label{eq:dadt}
\end{eqnarray}
After solving for $\ta(t)$, the output field is then found by using the input-output relation $\tao = \sqrt{2\kappa}\ta - \tai$.

The mechanical position and momentum quadratures are $\XM = (\b + \bd)/\sqrt{2}$ and $\PM = i(\bd - \b)/\sqrt{2}$, respectively, the cavity  (and its input/output) quadratures  are similarly defined via $\ta$ ($\tai$/$\tao$). The statistics of the optical amplitude quadrature is unaffected by the interaction, however, the phase quadrature contains the phase dependent upon the mechanical position. It has output emerging from the cavity $\PLout(t) = \frac{g_0}{\kappa} \sqrt{N_p} \, \varphi(t) \XMin + 2\kappa e^{-\kappa t}\integlim{-\infty}{t}{t'} e^{\kappa t'} \PLin(t') - \PLin(t)$, where $\varphi(t) = (2\kappa)^{\frac{3}{2}} e^{-\kappa t}\integlim{-\infty}{t}{t'} e^{\kappa t'} \alpha(t')$ describes the accumulation of phase, $\XMin$ is the mechanical position prior to the interaction and the last two terms are the input phase noise contributions. $\PLout$ is measured via homodyne detection, i.e. $\PL = \sqrt{2} \integ{t} \alphalo(t) \PLout(t)$. To maximize the measure of the mechanical position the local oscillator envelope is chosen as $\alphalo(t) = \mathcal{N}_{\varphi} \, \varphi(t)$, where $\N_{\varphi}$ ensures normalization. The contribution of $\XMin$ in $\PL$ scales with $\chi = \sqrt{2} \frac{1}{\mathcal{N}_{\varphi}} \frac{g_0}{\kappa} \sqrt{N_p}$, which quantifies the mechanical position measurement strength. The mean and variance of $\PL$ are given in Eq.~(\ref{eq:meanVarPL}) for pure Gaussian optical input and together with $\Omega$ and (\ref{eq:propPL}) are used to determine $\Ups$, as given in (\ref{eq:Upsilon}). We have thus arrived, for our physical setting, at an operator which is known from generalised linear measurement theory (see for example~\cite{Caves1987}). Also, we note that (\ref{eq:Upsilon}) is equivalent to $\Ups = e^{i\Omega \XM} \bra{\PLm} e^{i \chi \XL \XM } \ket{0}$, though the non-unitary process of cavity filling and decay is not explicit. We also remark that the construction of $\Ups$ can be readily generalized to include non-Gaussian operations. 

The maximum $\chi$ is obtained for the input drive $\alphai(t) = \sqrt{\kappa} e^{-\kappa \abs{t}}$. This can be seen by noting that $\N^{-2}_\varphi = \integ{t} \varphi^2(t)$, which in Fourier space is $\N^{-2}_\varphi \propto \integ{\omega} (\omega^2 + \kappa^2)^{-2}\abs{\alphai(\omega)}^2$. Hence, for such cavity-based measurement schemes, the optimal drive has Lorentzian spectrum.  This drive, $\alpha(t)$ obtained from (\ref{eq:dalphadt}) and the local oscillator are shown in Fig. \ref{Fig:Setup}(b). The resulting optimal measurement strength is given by 
\begin{equation}
\chi = 2 \sqrt{5} \, \frac{g_0}{\kappa} \sqrt{N_p},
\label{eq:maxChi}
\end{equation}
and the mean momentum transfer due to $\alpha^2$ is $\Omega = \frac{3}{\sqrt{2}} \frac{g_0}{\kappa} N_p$.

We note that this optimization of the driving field may also be applied to cavity-enhanced pulsed measurement of the spin of an atomic ensemble~\cite{ref:spinSqueeze, ref:Teper2008} or the coordinate of a trapped ion/particle~\cite{ref:Trapping}. Particularly in the latter case, this will broaden the repertoire of measurement techniques available and may lead to some interesting applications.

\section{Finite mechanical evolution during interaction}
\label{app:FiniteInt}

In the model used above we have assumed that the mechanical position remains constant during the pulsed optomechanical interaction. Including finite mechanical evolution, the intracavity field dynamics (\ref{eq:dadt}) must be determined simultaneously with the mechanical dynamics. In the mechanical rotating frame with the conjugate quadratures $\XMBar, \PMBar$ these dynamics are solved to first order in $\omegam/\kappa$ resulting in the input-output relations:
\begin{equation}
\begin{split}
\PMBarout & = \PMBarin + \Omega + \N_1 \chi \XC{1},\\
\XMBarout & = \XMBarin - \frac{\omegam}{\kappa} \xi_1 \Omega - \frac{\omegam}{\kappa} \chi \N_2 \XC{2},\\
\PL & = \PLin + \chi \left(\XMBarin + \frac{\omegam}{\kappa} \xi_2 \PMBarin \right) \\ & \quad +
\chi \frac{\omegam}{\kappa} \xi_3 \Omega + \chi^2 \frac{\omegam}{\kappa} \N_3 \XC{3},
\end{split}
\end{equation}
where $\PL$ still represents the measurement outcome, $\N_{1,2,3}$ and $\xi_{1,2,3}$ are input drive dependent dimensionless parameters of order unity, the former normalizing the non-orthogonal amplitude quadrature temporal modes $\XC{1,2,3}$. The main effects of the finite mechanical evolution can be seen in $\PL$. i) The mechanical quadrature measured has been rotated, which in terms of the non-rotating quadratures is $\XMTilde \simeq \XM + \frac{\omegam}{\kappa} \xi_2 \PM$. Such a rotation poses no principle limitation to our scheme however this must be taken into account for the measurement of a particular mechanical quadrature. ii) Each pulsed measurement now has a non-zero mean proportional to $\Omega$. This can be experimentally characterized and appropriately subtracted from the outcomes. iii) $\PL$ now includes a term proportional to the optical amplitude noise. This term decreases the back-action evading quality of the measurement and has arisen due to mechanical momentum noise gained from the optical amplitude quadrature evolving into position noise. The conditional variance of the rotated mechanical quadrature including these effects, for large initial occupation, is
\begin{equation}
\var{\XMTilde} \simeq \frac{1}{2}\left[ \frac{1}{\chi^2} + \zeta^2 \chi^2 \left(\frac{\omegam}{\kappa}\right)^2 \right],
\end{equation}
where $\zeta$ is another drive-dependent parameter of order unity. The two competing terms here give rise to a minimum variance of $\zeta \omegam / \kappa$ when $\chi^2 = \kappa / (\zeta \omegam)$. Experimentally reasonable values of $\chi$ will lie much below this optimum point, however, as $\kappa \gg \omegam$ for the parameters we consider, the broadening due to finite evolution is small and strong squeezing can be achieved.

%As $\kappa \gg \omegam$ this minimal variance is  highly squeezed and experimentally reasonable values of $\chi$ will lie much below this optimum point.

\begin{acknowledgments}
We thank the ARC (grant FF0776191), UKEPSRC, ERC (StG QOM), European Commission (MINOS, Q-ESSENCE), FQXi, FWF (L426, P19570, SFB FoQuS, START), QUEST, and an \"{O}AD/MNiSW program for support. M.R.V. and I.P. are members of the FWF Doctoral Programme CoQuS (W 1210). M.R.V. is a recipient of a DOC fellowship of the
Austrian Academy of Sciences. G.D.C. is a recipient of a Marie Curie Fellowship of the European Commission. The kind hospitality provided by the University of Gda\'{n}sk (I.P.), the University of Queensland (M.R.V.) and the University of Vienna (G.J.M.) is acknowledged. We thank S. Hofer, N. Kiesel and M. \.{Z}ukowski for useful discussion.
\end{acknowledgments}

\end{document}